\documentclass[final,3p,times]{elsarticle}
\usepackage{amssymb}
\usepackage{amsmath}
\begin{document}

\begin{frontmatter}
\title{An \emph{ab initio} study of structural phase transitions of crystalline aluminum under ultrahigh pressures based on ensemble theory}
\author[1,2,3]{Bo-Yuan Ning\corref{cor1}}
\ead{byning@fudan.edu.cn}
\author[4]{Li-Yuan Zhang\fnref{fn1}}
\cortext[cor1]{To whom the correspondence should be addressed.}
\fntext[fn1]{Present address: Institute of Modern Physics, Fudan University, Shanghai, 200433, China}
\affiliation[1]{organization={Institute of Modern Physics, Fudan University},%Department and Organization
  city={Shanghai},
  postcode={200433}, 
  country={China}}
\affiliation[2]{organization={Applied Ion Beam Physics Laboratory, Fudan University},%Department and Organization
  city={Shanghai},
  postcode={200433}, 
  country={China}}
\affiliation[3]{organization={Department of Materials Science and Engineering, Southern University of Science and Technology},%Department and Organization
  city={Shenzhen},
  postcode={518055}, 
  country={China}}
\affiliation[4]{organization={School of Physics, East China University of Science and Technology},%Department and Organization
  city={Shanghai},
  postcode={200237}, 
  country={China}}

\begin{abstract}
It is a long-time pursuit of computations with \emph{ab initio} precision of thermal contributions to phase behaviors of condensed matters under extreme conditions. In this work, the pressure induced structural phase transitions of crystalline aluminum up to $600$ GPa at room temperature are investigated based on the criterion of Gibbs free energy derived directly from the partition function that formulated in the ensemble theory with the interatomic interactions characterized by density functional theory computations. The transition pressures of the FCC$\rightarrow$HCP$\rightarrow$BCC phase transitions are determined at $194$ and $361$ GPa, the axial ratio of the stable HCP structure is found to be equal to $1.62$ and the discontinuities in the equations of states are confirmed to be associated with $-0.67\%$ and $-0.90\%$ volume changes, which are all in an excellent agreement with the measurements by one of the recent experiments but differ from other experimental observations. Compared with the results obtained by the criterion of enthalpy at $0$K, this work further shows the nontrivial thermal impacts on the structural stability of aluminum under ultrahigh-pressure circumstances even at room temperature.
\end{abstract}

% \includegraphics[width=4.5in,height=3.5in]{abstract-img}

% \begin{highlights}
% \item An investigation on structural phase transitions of crystalline Al from the perspective of statistical ensemble theory, which is different from all the previous theoretical works.
% \item Implementation of a newly develop method, direct integral approach, to solve partition functions, Gibbs free energy and equations of state parameter-freely with \emph{ab initio} precision.
% \item The theoretically obtained results up to $600$GPa confirm the measurements from one of the latest experiments [Nature Communications \textbf{9} 2913 (2018)], and settle down the disputes over the values of two transition pressures.
% \item Thermal contributions to the Gibbs free energy at room temperature are quantitatively computed and are shown with nontrivial effects.
% \end{highlights}


\end{frontmatter}


\section{Introduction}
\label{sec:1}
The rapid development of experimental techniques makes it possible to explore the phase behaviors of condensed matters under extreme conditions\cite{hp1,hp2} and simultaneously requests more substantial understandings from the perspective of theoretical physics.
Aluminum (Al), as one of the most important metallic materials for various industrial applications,
serves as an ideal benchmark system for high-pressure studies due to its simple $sp$-electron shells\cite{pickard2010},
and the structural phase transitions of crystalline Al under ultrahigh pressures attract extensive attentions from both experimental and theoretical works.

Although the face-centered-cubic (FCC) structure was believed to be the stable phase up to $1000$ GPa at room temperature according to early experiments\cite{al1988,al1994,dewaele2004} and \emph{ab initio} computations\cite{afe2000},
following experiments of static compressions\cite{akahama2006,guillaume2019} and laser-driven dynamic ramp compressions\cite{polsin2017} observed a sequential structural transition of FCC$\rightarrow$hexagonal-close-packed (HCP)$\rightarrow$body-centered-cubic (BCC)
and the two transition pressures were determined to be about $220$ and $320$ GPa,
which distinctly differ from the recently measured $198$ and $360$ GPa by Dewaele et al.\cite{Dewaele2018} who attributed the differences to be the usage of biased empirical equation of state (EOS) of the pressure markers in previous works or the difficulty in controlling the temperature effects on the ramp compressions.

The discrepancy of transition pressures remains to exist among theoretical investigations\cite{boettger1996,smirnov2002,marcus2006,tambe2008,mubarak2020} in spite of the agreements on the transition sequence.
By using the criterion of enthalpy at $0$K obtained by density functional theory (DFT) computations, 
Boettger et al.\cite{boettger1996} predicted the two transition pressures to be $205$ and $565$ GPa,
while the results of Sinrquotko et al.\cite{smirnov2002} are about $165$ and $381$ GPa,
which also differ from the predicted $192$ GPa for the FCC$\rightarrow$HCP transition by Jona et al.\cite{marcus2006} but are close to the $175$ and $380$ GPa reported by Tambe et al.\cite{tambe2008} and Eidelstein et al.\cite{mubarak2020}.
As noted by Tambe et al.\cite{tambe2008},
the differences of transition pressures between experiments and theoretical models are likely to stem from the exclusion of lattice thermal contributions, while the transition pressure of FCC$\rightarrow$HCP was reported to be $290$ GPa by Boettger et al.\cite{boettger1996} conducting a rough estimation of phonon vibrations and Sinrquotko et al.\cite{smirnov2002} predicted the two pressures to be $232$ and $277$ GPa by performing phonon calculations based on quasi-harmonic approximation,
the deviations of which from experimental measurements can be understandable because the accuracy of the calculated entropy either by phonon spectrum or the empirical Debye model was recently questioned by Argaman et al.\cite{argman2015} as well as considering other unsatisfactory results regarding the thermal expansion coefficient or conductivity\cite{phonon2003,phonon2005,phonon2018,phonon2019}.
Due to the challenge in face of \emph{ab initio} computations to include finite-temperature effects\cite{liu2020},
in terms of crystalline Al, the most precise pathway so far was resorted to DFT computations combined with the semi-empirical SESAME EOS\cite{sjostrom2016}
and the two transition pressures were determined at $185$ and $389$ GPa respectively,
which deviate about $15\%$ from the measurements in Refs.\cite{polsin2017, guillaume2019} and about $7\%$ from those in Ref.\cite{Dewaele2018}.

To settle down the above disputes over the values of the transition pressures,
the key point is to accurately compute the finite-temperature free energy (FE) that governs the phase stability,
and the statistical ensemble theory has already paved a stringent way that 
all the thermodynamic state functions can be readily obtained parameter-freely as long as the partition function (PF) is solved\cite{fe1,pes1}.
Nevertheless, 
the exact solution to the complex high-dimension configurational integral in PF of condensed matters is a long-standing problem so that
certain serious concerns were raised on whether the orthodox ensemble theory is capable of correctly describing the first-order phase transitions of condensed matters\cite{pfconden,mce}.
With a nearly century-long effort,
many algorithms to PF have been put forward\cite{ce2020,metadynamics3} and
the Monte-Carlo-based nested sampling\cite{nestsample3,do4,nestsample13} may be state-of-the-art one that greatly improved the computational precision and efficiency compared with previous ones such as thermodynamics integration\cite{ti2} or Wang-Landau sampling\cite{wl1}.
Unfortunately,
for tackling with realistic systems that demand precisely characterized interatomic interactions,
especially for high pressure-temperature cases,
the efficiency of the sampling algorithm is still not high enough to enable the fast developing \emph{ab initio} computations\cite{dft1,dft5,dft3,dft4} but limits to the usage of empirical potentials,
and the two transition pressures of crystalline Al obtained by the nested sampling are $\sim20$ and $\sim50$ GPa respectively when trying different empirical embedded-atom-method potentials\cite{nspv}.
Because of the dilemma of solving PF, 
more researches on finite-temperature phase behaviors of metal materials turned to the model of approximated FE without the knowledge of PF\cite{lattice2012,afe2019,afe2004,afe2002,afe1993,latticedynamics,jha2006},
which has been proved to produce valuable results but may not substantially address all the issues.

Very recently, we put forward a direct integral approach (DIA) to the PF of condensed matters with ultrahigh efficiency and precision\cite{nby,lyp,glc1,glc2}
which successfully incorporated the DFT computations into studying the finite-temperature phase transitions of vanadium\cite{nby2022}, the EOS of copper\cite{nby} and the optimum growth condition for $2$-dimension materials\cite{lyp}.
Compared with quasi-harmonic phonon model,
DIA was examined to be applicable to a much wider realm with a much higher precision\cite{glc2}.
The aim of this work is to apply DIA to compute the PFs of crystalline Al
and use the Gibbs FE as the criterion to carefully investigate the structural phase transitions up to $600$ GPa as well as examine the usually overlooked thermal impacts on the phase stability at a room-temperature condition.
The paper is organized as follows:
In Sec.\ref{sec:2}, the theoretical model of DIA to PF is briefly reviewed,
and, in Sec.\ref{sec:3}, the detailed implementations of DIA to crystalline Al as well as key parameters of DFT computations are presented.
The phase transitions determined by the criterion of enthalpy at $0$K and of Gibbs FE at $300$K are discussed and compared in Sec.\ref{sec:4},
and finally,
a conclusion is made in Sec.\ref{sec:5}.

% Although the quantum mechanics \emph{ab initio} simulations paved a promising way for researches on the properties from the atomic scale\cite{dft1,dft5,dft3,dft4}
% and progress has been achieved in predicting the structural stability by the global optimization algorithms\cite{glopt4,glopt2,glopt3,glopt5,glopt1}
% that commonly adopt the criteria of potential energy or enthalpy obtained by electronic-structure computations within the Born-Oppenheimer approximation at $T=0$K,
% Based on \emph{ab initio} computations,
% the phonon model within the quasi-harmonic approximation is widely used in order to include the thermal contributions to the FE\cite{afe2002,afe2019,afe2005,afe2004,afe1993}
\section{Theoretical Model of DIA}
\label{sec:2}
The PF defined in the ensemble theory for a system consists of $N$ particles with their Cartesian coordinate $\mathbf{q}^N=\{\mathbf q_1,\mathbf q_2,\ldots\mathbf q_N\}$ confined within a volume $V$ at temperature $T$ reads
\begin{eqnarray}
  \label{eq:1}
  \mathcal{Z}&=&\frac{1}{N!}\left(\frac{2\pi m}{\beta h^2}\right)^{\frac{3}{2}N}\int d\textbf{q}^N\exp[-\beta U(\textbf{q}^N)] \nonumber \\
             &=&\frac{1}{N!}\left(\frac{2\pi m}{\beta h^2}\right)^{\frac{3}{2}N}\mathcal Q,
\end{eqnarray}
where $h$ is the Planck constant, $m$ is the particle mass, $\beta=1/k_BT$ with $k_B$ is the Boltzmann constant,
$U(\textbf{q}^N)$ is the total potential energy,
and $\mathcal Q=\int d\textbf{q}^N\exp[-\beta U(\textbf{q}^N)]$ is the so-called configurational integral
that is related to the structures of the system at given conditions.
If the configurational integral is solved,
then the pressure-volume ($P\text{-}V$) EOS and the Gibbs FE ($\mathcal G$) can be computed as
\begin{eqnarray}
  \label{eq:2}
  P&=&\frac{1}{\beta}\frac{\partial\ln\mathcal{Q}}{\partial V}, \\
  \label{eq:3}
  \mathcal G&=&-\frac{1}{\beta}\ln[\frac{1}{N!}\left(\frac{2\pi m}{\beta h^2}\right)^{\frac{3}{2}N}]-\frac{1}{\beta}\ln\mathcal{Q}+PV.
\end{eqnarray}
For crystalline systems with atoms locating at lattice sites, $\mathbf Q^N$, and
with the corresponding total potential energy, $U_0(\mathbf{Q}^N)$,
the DIA firstly introduces a transformation as
\begin{equation}
  \label{eq:4}
  \mathbf q'^N=\mathbf q^N-\mathbf Q^N,\ U'(\mathbf q'^N)=U(\mathbf q'^N)-U_0(\mathbf Q^N), 
\end{equation}
where $\mathbf q'^N$ and $U'(\mathbf q'^N)$ represent the displacements of atoms away from their lattice sites
and the corresponding differences of total potential energy with respect to the $U_0(\mathbf Q^N)$,
and the configurational integral is thus expressed as
\begin{equation}
  \label{eq:4a}
  \mathcal Q=e^{-\beta U_0(\mathbf{Q}^N)}\int e^{-\beta U'(\mathbf q'^N)}d\mathbf q'^N.
\end{equation}
By our introduced reinterpretations of integrals\cite{nby},
the $3N$-fold integral in Eq.(\ref{eq:4a}) is mapped to an effective $3N$-dimensional volume,
and the value of the volume may be approximated as 
\begin{equation}
  \label{eq:4b}
  \mathcal Q=e^{-\beta U_0(\mathbf Q^N)}\prod_{i=1}^{N}\mathcal L_{i_{x}}\mathcal L_{i_{y}}\mathcal L_{i_{z}}
\end{equation}
where $\mathcal L_{i_{x(,y,z)}}$ is called the effective length of $i$th atom along $x$ ($y$ or $z$) axis and defined as
\begin{equation}
  \label{eq:4c}
  \mathcal L_{i_{x(,y,z)}}=\int e^{-\beta U'(q'_{i_{x(,y,z)}})}dq'_{i_{x(,y,z)}},
\end{equation}
where the potential-energy curve (function), $U'(q'_{i_{x(,y,z)}})$,
is obtained by moving the $i$th atom along the $x$ ($y$ or $z$) axis
while the other two degrees of freedom of the atom and all the other atoms are kept fixed.

For pure-element systems with FCC or BCC structures,
as shown in our previous works for copper\cite{nby} and vanadium\cite{nby2022} respectively,
since all the atoms in the lattice are geometrically equivalent and
the potential-energy curve $U'_x$ felt by an arbitrary atom moving along $x$ axis is the same as the one
along $y$ or $z$ axis,
their configurational integrals in Eq.(\ref{eq:4b}) thus can be further simplified as
\begin{equation}
  \label{eq:5}
  \mathcal{Q}=e^{-\beta U_0(\mathbf Q^N)}\mathcal L^{3N},
\end{equation}
where $\mathcal L$ represents the effective length of an arbitrary atom along either $x$, $y$ or $z$ axis. 
In terms of the HCP structure,
although all the atoms are also geometrically equivalent,
the effective lengths along different axes of an atom are not equivalent,
and consequently, all the three effective lengths along $x$, $y$ and $z$ axes of an arbitrary atom have to be considered for the computations of the configurational integral as
\begin{equation}
  \label{eq:6}
  \mathcal{Q}=e^{-\beta U_0(\mathbf Q^N)}(\mathcal L_x\mathcal L_y\mathcal L_z)^{N}.
\end{equation}
For DIA to be applied to more complex structures or multi-element systems,
it is needed to make more classifications of groups on the basis of the specific lattice geometry and
we refer to our previous works of large molecules\cite{nby} and $2$-D materials\cite{lyp} for details.

\begin{figure}
  \includegraphics[width=2.4in, height=3.4in]{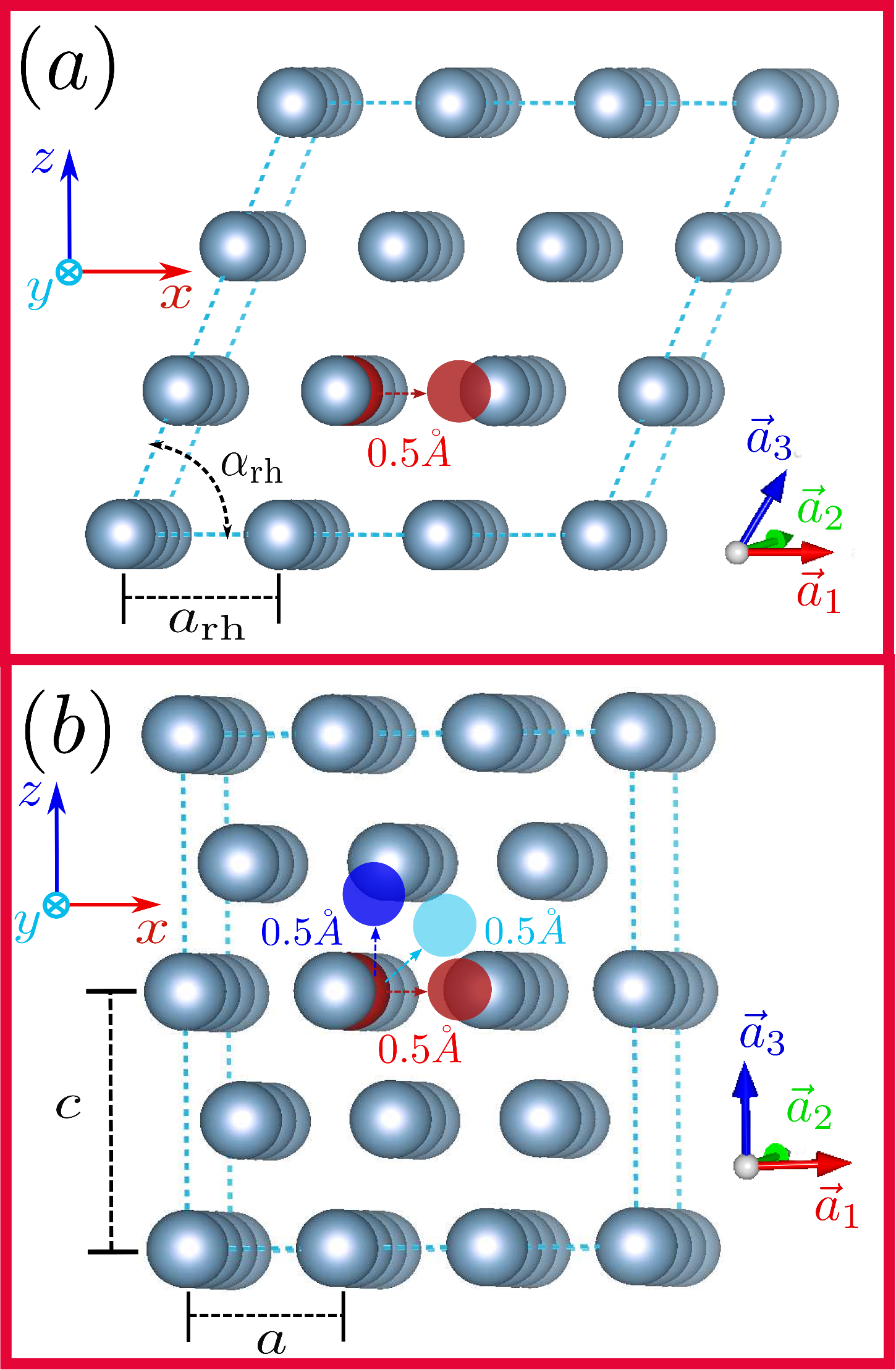}
  \centering
\caption{\label{fig:1}(Color online) Schematics of implementations of DIA. (a) The FCC and BCC structures are placed in a rhombohedral lattice with lattice parameter of $a_{rh}$ and rhombohedral angle $\alpha_{rh}$, and an arbitrarily selected atom is moved from its lattice site (colored in opaque red) along the $x$ axis by $0.5$ {\AA} (colored in transparent red) for the computation of effective length $\mathcal L$ defined in Eq.(\ref{eq:5}). (b) The HCP structure is placed in a hexagonal lattice with lattice parameter of $a$ and axial ratio $c/a$, and an arbitrarily selected atom is moved from its lattice site (colored in opaque red) along the $x$, $y$ and $z$ axis separately by $0.5$ {\AA} (colored in transparent red, cyan and blue respectively) for the computations of effective lengths $\mathcal L_x$, $\mathcal L_z$ and $\mathcal L_z$ defined in Eq.(\ref{eq:6}). The view angles of the shown lattices are slightly tilted for a more clear illustration.}
\end{figure}

At first glance,
the model of DIA may be argued to be similar to the phonon model that is based on the assumption of quasi-harmonic motions of atoms around their lattice sites,
while, as can be seen in Eqs.(\ref{eq:4b}) and (\ref{eq:4c}),
the key difference lies in that
the DIA has no particular restrictions on the motions of thermal atoms, 
the contributions of which to PF only depend the value of the effective length that is determined by temperature (or Boltzmann factor $e^{-\beta U'_{i_{x(y,z)}}}$) and 
the potential-energy curves $U'_{i_{x(y,z)}}$ that are decided by the intrinsic properties of the system,
e.g., lattice parameters, lattice symmetry, strength of interatomic interactions etc..
In other words,
the thermal contributions from both harmonic and anharmonic lattice motions are naturally included in the model of DIA
and our previous simulations have proved the accuracy of DIA over the model of quasiharmonic approximation\cite{glc2}.

\begin{table*}
\caption{\label{tab1} Transition pressures for crystalline Al at $300$K determined by theoretical and  experimental works.}
% \begin{ruledtabular}
\centering
\begin{tabular}{c|cc|cc|cccc}
\cline{1-9}
 &\multicolumn{4}{c}{Theoretical Results (GPa)}&\multicolumn{4}{c}{Experimental Results (GPa)}\\ \hline
 &\multicolumn{2}{c}{Present Work } & Ref.\cite{smirnov2002}& Ref.\cite{sjostrom2016} & Ref.\cite{akahama2006} & Ref.\cite{polsin2017} & Ref.\cite{guillaume2019} & Ref.\cite{Dewaele2018}\\ \hline
  method & DIA & $0$K-enthalpy & \emph{ab initio} phonon & \emph{ab initio} + EOS & DAC & dynamic compression & DAC & DAC \\ \hline
  FCC$\rightarrow$HCP & 194 & 171 & 232 & 185 & 217 & 216 & 220 & 198 \\ \hline
  HCP$\rightarrow$BCC & 361 & 378 & 277 & 389 & / & 321 & 320 & 360 \\
\cline{1-9}
\end{tabular}
% \end{ruledtabular}
\end{table*}

\section{\label{sec:3}Computational Details}
\subsection{\label{sec:3:1} Implementations of DIA}
As shown in Fig.\ref{fig:1}(a),
the FCC and BCC structures of crystalline Al,
for the sake of less computational cost,
are placed in a $3\times3\times3$ rhombohedral supercell with one atom in the unit cell where the lattice vectors are set as
$\mathbf{a}_1=a_{\text{rh}}\cdot(1, 0, 0)$, $\mathbf{a}_2=a_{\text{rh}}\cdot(\cos\alpha_{\text{rh}}, \sin\alpha_{\text{rh}}, 0)$, $\mathbf{a}_3=a_{\text{rh}}\cdot(\cos\alpha_{\text{rh}}, \frac{\cos\alpha_{\text{rh}}-\cos^2\alpha_{\text{rh}}}{\sin\alpha_{\text{rh}}}, \frac{\sqrt{1-3\cos^2\alpha_{\text{rh}}+2\cos^3\alpha_{\text{rh}}}}{\sin\alpha_{\text{rh}}})$ with $a_{\text{rh}}$ denoting the lattice parameter and the rhombohedral angle $\alpha_{\text{rh}}=60^\circ$ and $109.47^\circ$ for FCC and BCC respectively. 
The HCP structure is placed in a $3\times3\times2$ typical hexagonal supercell with two atoms in the unit cell where the lattice vectors are set as
$\mathbf{a}_1=a\cdot(1, 0, 0)$, $\mathbf{a}_2=a\cdot(\frac{1}{2}, \frac{\sqrt{3}}{2}, 0)$, $\mathbf{a}_3=a\cdot(0, 0, c/a)$ with $a$ and $c/a$ standing for the lattice parameter and axial ratio respectively as shown in Fig.\ref{fig:1}(b).
To apply the DIA,
the selected atom (colored in opaque red in Fig.\ref{fig:1}) in all the three structures are moved by $0.5$ {\AA} at a step of $0.05$ {\AA} along $x$ (for FCC, BCC) or $x$, $y$, $z$ axes (for HCP).
The potential energies at each step are calculated by the DFT computations and
afterwards smoothened by the spline interpolation algorithm\cite{spl1,spl2} to obtain the potential-energy curves (see Tables.S1-S7 and Figs.S1-S7 in the supplementary materials).

\subsection{\label{sec:3:2} Details of DFT Computations}
All the DFT computations are performed in Vienna Ab initio Simulation Package\cite{VASP1,VASP2} where the projector-augmented wave formalism\cite{paw1,paw2} with $3$ valence electrons ($3s^2p^1$) considered is used for the pseudopotential and the general gradient approximation of the Perdew-Burke-Ernzerhf parametrizations\cite{pbe} is adopted for the exchange-correlation functional of electrons.
% which, in principle, may be replaced by the fast-developing finite-temperature functionals\cite{tefunc2014,tefunc2015,tefunc2017,tefunc2018,tefunc2021} to achieve possibly higher computational precision in future works.
The cut-off energy of the plane-wave basis is set as $312.4$ eV and the Brillouin zone is sampled by a $\Gamma$-centered $9\times9\times9$ uniform $k$-mesh grid using the Monkhorst-Pack scheme\cite{monkhorst} to guarantee that the mesh spacing at least equals to or is better than $2\pi\times0.015$ {\AA}$^{-1}$ for all the three structures with different atomic volumes.
The tetrahedron method with Bl\"ochl correction is used to handle the fractional occupancies of electron orbitals and the stop condition of the electronic self-consistent calculations for total potential energy is set as $10^{-6}$ eV.

\begin{figure}
\includegraphics[width=3.3in, height=2.5in]{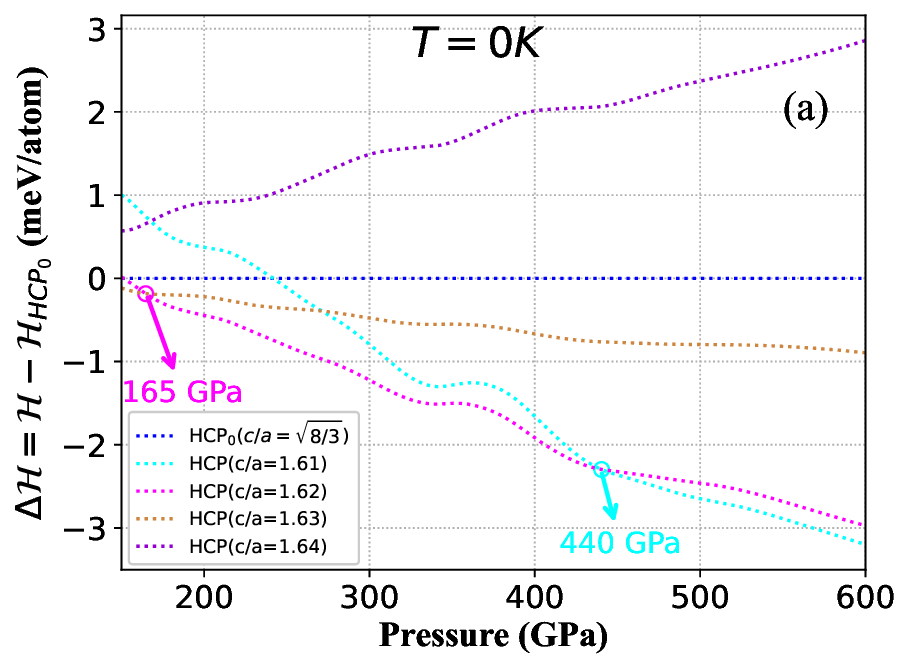}
\includegraphics[width=3.3in, height=2.5in]{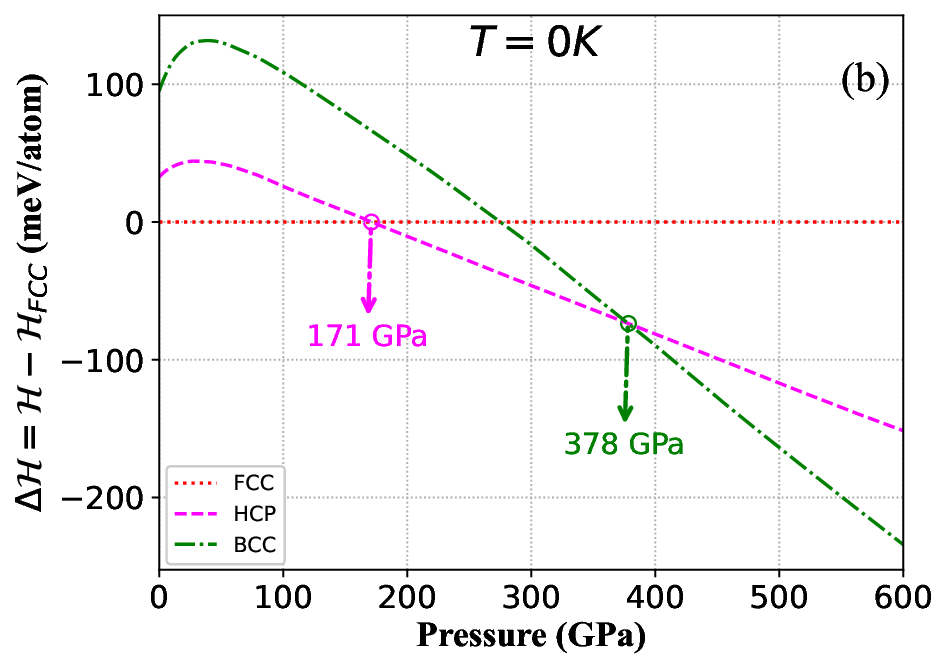}
\caption{\label{fig:2}(Color online) Structural phase transitions of crystalline Al under high pressures determined by the criterion of enthalpy at $0$K. (a) The enthalpy differences of four HCP structures with $c/a=1.61$, $1.62$, $1.63$ and $1.64$ (denoted by cyan, magenta, brown and violet dotted lines respectively) relative to the one with ideal axial ratio of $c/a=\sqrt{8/3}$ (blue dotted line) in a pressure range from $150$ to $600$ GPa. (b) The enthalpy differences of HCP and BCC structures (denoted by magenta dashed line and green dash-dotted line respectively) relative to that of FCC structure (red dotted line) in a pressure range from $0$ to $600$ GPa.}
\end{figure}

\section{\label{sec:4}Results and Discussions}
\subsection{\label{sec:4:1} Phase transitions at $0$K}
We first consider the phase transitions at $0$K and used the criterion of enthalpy,
which is defined as $\displaystyle\mathcal H=U_0+PV$ with $U_0$ being the total potential energy and the pressure being calculated as $P=\partial U_0/\partial V$. 
At high-pressure zone,
since the optimum axial ratio of the HCP structure may not equal to the ideal value as $c/a=\sqrt{8/3}$,
which was also noticed in Refs.\cite{tambe2008,marcus2006},
besides the one with the ideal axial ratio, 
we consider additional four HCP structures with the $c/a$ values of $1.61$, $1.62$, $1.63$, and $1.64$,
and compute their enthalpies from $150$ to $600$ GPa,
in the pressure range of which the structural transitions of crystalline Al are likely to take place.
Fig.\ref{fig:2}(a) shows the enthalpy differences of the four HCP structures relative to the one with ideal axial ratio,  
and the two cross points indicate that 
the axial ratio of the stable HCP structure be $1.62$ instead of the ideal value when the pressure reaches $165$-$440$ GPa, and then, turn to be $1.61$ up to $600$ GPa.
The obtained axial ratio of the HCP phase is slightly different from previous works where $c/a$ varies from $1.63$ to $1.61$ in a range of $150$-$450$ GPa\cite{tambe2008} or is around $1.63$ within $180$-$340$ GPa\cite{marcus2006},
but agrees well with the experimental value measured in Refs.\cite{akahama2006,Dewaele2018}.

With the determined dependence of the HCP axial ratio on pressures,
the enthalpies of FCC, HCP and BCC structures in a pressure range from $0$ to $600$ GPa are computed, and
the enthalpy differences relative to that of the FCC are shown in Fig.\ref{fig:2}(b).
The two cross points of the enthalpy curves in Fig.\ref{fig:2}(b) clearly identify the sequential structural phase transitions of FCC$\rightarrow$HCP$\rightarrow$BCC 
and the transition pressures locate at $171$ and $378$ GPa respectively,
the value of which coincide with those reported in Refs.\cite{tambe2008,mubarak2020} who used the same frozen-core pseudopotential method of DFT computations
and confirm the accuracy of our DFT computations.
On the other hand,
when compared with experiments,
the two transition pressures determined at $0$K are not satisfactory because the pressure values deviate $\sim20\%$ from the experimental results in Refs.\cite{polsin2017,guillaume2019} and 
$\sim10\%$ from those in Ref.\cite{Dewaele2018},
which implies the importance of thermal contributions,
especially in a consideration of the more accurate results in Ref.\cite{sjostrom2016}
where the thermal contribution is included to a certain extent by utilizing the empirical EOS.

\begin{figure}
  \centering
  \includegraphics[width=3.2in, height=2.4in]{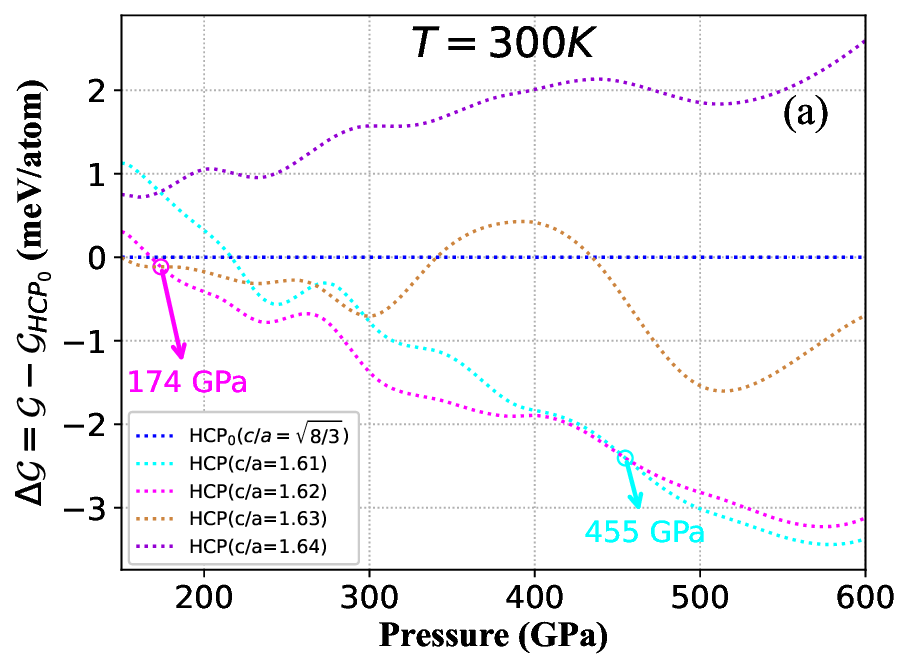} 
  \includegraphics[width=3.2in, height=2.4in]{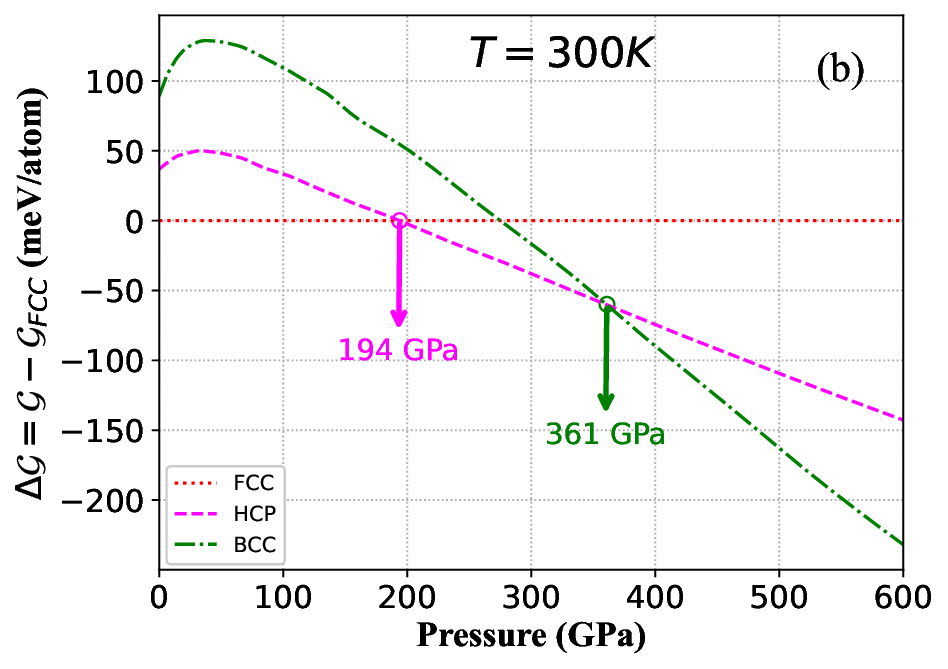} 
\caption{\label{fig:3} Similar to Fig.\ref{fig:2} except that the structural phase transitions are determined by the criterion of Gibbs FE at $300$K.}
\end{figure}

\subsection{\label{sec:4:2} Phase transitions at $300$K}
We next apply DIA to calculate the Gibbs FE as the criterion to investigate the phase transitions at $300$K and 
the first step is also to determine the optimum axial ratio among the five HCP structures within the pressure range of $150$-$600$ GPa.
As shown in the Fig.\ref{fig:3}(a),
the HCP structure with axial ratio value of $1.62$ would become the stable one when the pressure arrives at $174$ GPa
and be replaced by that of $1.61$ at $455$ GPa,
the transition order of which is the same as that obtained by the criterion of enthalpy at $0$K
while the pressure range of which is lifted by about $10$ GPa.

After the determinations of HCP structures,
the Gibbs FEs of the FCC, BCC and HCP structures from $0$ to $600$ GPa at $300$K are calculated by DIA according to Eqs.(\ref{eq:5}) and (\ref{eq:6}) respectively, 
and the differences relative to the FCC are shown in Fig.\ref{fig:3}(b). 
The structural phase transitions of FCC$\rightarrow$HCP$\rightarrow$BCC is reproduced that 
the stable phase of the minimum Gibbs FE remains to be the FCC until the pressure increases to $194$ GPa and then becomes to be the HCP with the axial ratio of $1.62$ until $361$ GPa,
from where the BCC finally replaces the HCP to be the stable one up to $600$ GPa.
For the transition pressure of FCC$\rightarrow$HCP at $300$K, 
the $194$ GPa determined by DIA is different from either the $232$ GPa in Ref.\cite{smirnov2002} based on phonon model or the $290$ GPa reported in Ref.\cite{boettger1996} with rough calculations of phonon contributions, but is close to the transition pressure of $185$ GPa from Ref.\cite{sjostrom2016} which relies on the empirical SESAME EOS.
For the transition pressure of HCP$\rightarrow$BCC at higher pressure zone, 
as listed in Table.\ref{tab1},
the difference between the result by DIA, $361$ GPa, and that by phonon model, $277$ GPa, becomes much more prominent while
the $389$ GPa obtained in Ref.\cite{sjostrom2016} is still close to our result.

Compared with the experimental values of the transition pressures in Refs.\cite{akahama2006,polsin2017,guillaume2019} (see Table.\ref{tab1}),
the relative deviations of DIA, $|P_{DIA}-P_{exp}|/P_{exp}$,
are about $11\%$ and $13\%$ for the FCC$\rightarrow$HCP and HCP$\rightarrow$BCC transitions respectively,
which are all better than the results from Ref.\cite{sjostrom2016}.
Furthermore,
the relative deviations of DIA is about $2\%$ and $0.3\%$ away from the measurements by one of the latest experiment by Dewaele et al.\cite{Dewaele2018},
which are by far the least differences between theoretical computations and experiments for crystalline Al to our best knowledge.
It should be noted here that
the axial ratio, $c/a$, of the HCP structure determined in Ref.\cite{Dewaele2018} is the same as that obtained by DIA, $1.62$,
while the value is $1.64$ measured in Refs.\cite{polsin2017,guillaume2019}.
In terms of our computations as shown in Fig.\ref{fig:3}(a),
however,
the influence from the axial ratio of the five HCP structures to the transition pressures is within $5$ GPa due to the Gibbs FE differences of the HCP structures being about $1\sim2$ meV/atom
and cannot simply explain the comparatively large deviations of DIA away from Refs.\cite{polsin2017,guillaume2019}.
Another potential factor may be from the intrinsic precision limit of the DFT computations that, 
according to our estimations (see in supplementary materials)
a possible variation of $1$ meV/atom for the total potential energies
would roughly result in $\sim 0.2$ GPa variations of pressure and $\sim 15$ meV/atom variations of Gibbs FE in the pressure range of $200$-$400$ GPa,
which may slightly affect the values of the obtained transition pressures by DIA and be further carefully scrutinized in future,
while the current computational precision is believed to be reliable enough to predict accurate results considering that
the thermal contributions are overall larger than $200$ meV/atom as discussed in the next subsection.

\begin{figure}
\includegraphics[width=3.4in, height=2.6in]{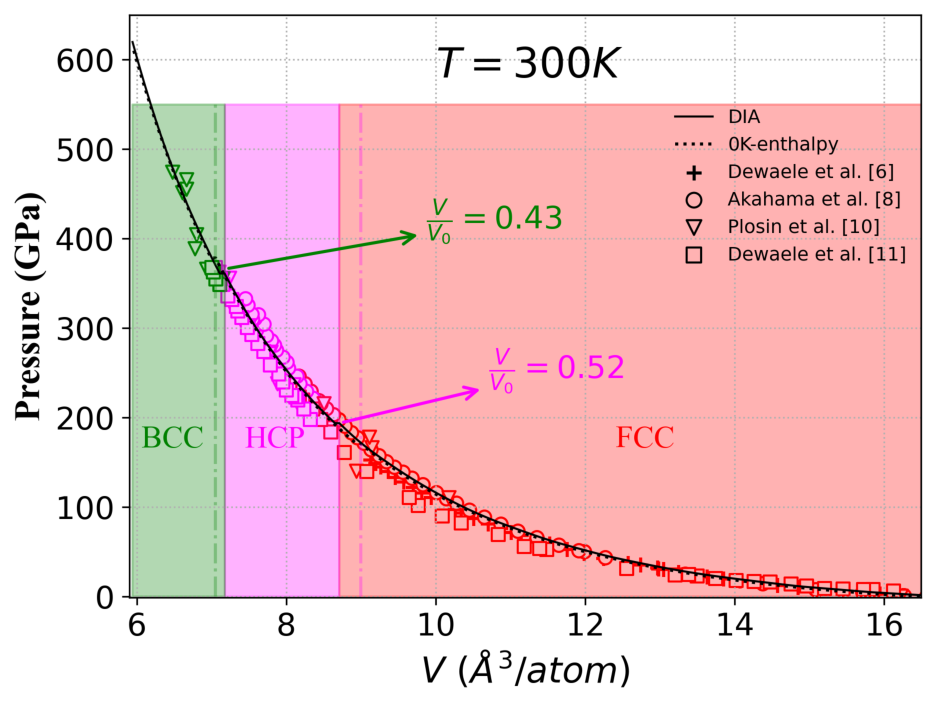} 
\caption{\label{fig:4}(Color online) Comparisons of the $P$-$V$ EOS at $T=300$K determined by DIA (black solid line), as well as the one at $0$K (black dotted line), with experimental data in Refs.\cite{dewaele2004} (crosses), \cite{akahama2006} (circles), \cite{polsin2017} (downward triangles) and \cite{Dewaele2018} (squares) with the FCC, HCP and BCC phases colored in red, magenta and green respectively. The two vertical dashed-dotted lines denote the phase boundaries of FCC$\rightarrow$HCP (colored in magenta) and HCP$\rightarrow$BCC (colored in green) determined by the criterion of enthalpy at $0$K.}
\end{figure}

With the transition pressures determined by Gibbs FE,
the $P$-$V$ EOS up to $600$ GPa at $300$K is solved by DIA according to Eq.(\ref{eq:2}).
As shown in Fig.\ref{fig:4},
our determined EOS agrees well with the existing experimental data and does not show much tendency towards any of the experiments,
which is shown by the comparisons that 
it is overall about $3\%$ lower than the EOS by Ref.\cite{akahama2006},
about $4\%$ higher the one by Ref.\cite{Dewaele2018} and
about $3\%$ different from the one by Ref.\cite{polsin2017}, and, 
the atomic volume of the FCC structure at standard conditions based on the EOS by DIA,
$V_0=16.845$ {\AA}$^3$/atom,
differs about $1.5\%$ from the values of $16.573$ {\AA}$^3$/atom in Refs.\cite{dewaele2004,Dewaele2018} and $16.597$ {\AA}$^3$/atom in Ref.\cite{akahama2006}.
On the other hand,
the phase boundaries of the three structures are determined to locate at $V/V_0=0.52$ for the FCC$\rightarrow$HCP and $V/V_0=0.43$ for the HCP$\rightarrow$BCC, 
accompanied with two prominent discontinuities of $-0.67\%$ and $-0.90\%$ volume changes at the transition points respectively,
which are very close to the reported volume changes of $-1.0\%$ and $-0.8\%$ in Ref.\cite{Dewaele2018},
but far less than the values of $-3.2\%$ and $-2.7\%$ observed in the Ref.\cite{polsin2017} using laser-driving dynamic compressions.
With the above comparisons and taking into account other theoretical results based on phonon model\cite{boettger1996,smirnov2002} or relied on empirical EOS\cite{sjostrom2016},
our current results are more inclined to support the experimental measurements by Dewaele et al.\cite{Dewaele2018}.

\begin{figure}
\includegraphics[width=3.4in, height=2.6in]{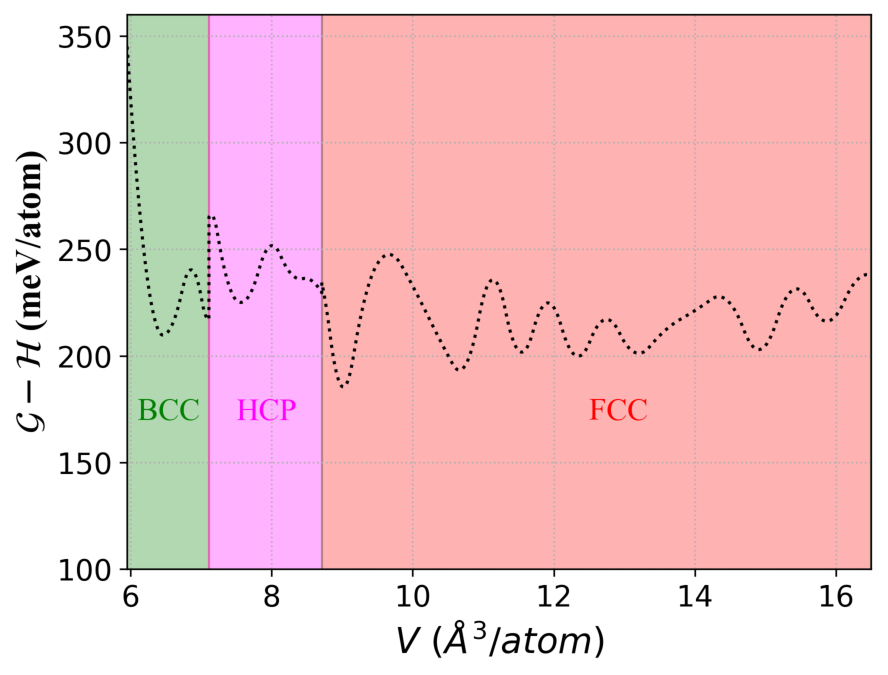} 
\caption{\label{fig:5}(Color online) The differences (black dotted line) between Gibbs FE at $300$K determined by DIA and enthalpy at $0$K with the FCC, HCP and BCC phases colored in red, magenta and green respectively.}
\end{figure}
\subsection{\label{sec4:3} The thermal contributions at $300$K}
Since there exist obvious differences of the transition pressures determined by DIA at $300$K and by the criterion of enthalpy at $0$K,
it is worthy further inspecting how much the thermal effects contribute to Gibbs FE of crystalline Al at room temperature.
The thermal contributions are obtained by the difference between Gibbs FE at $T=300$K and enthalpy as
\begin{equation}
\Delta_{thermal}=\mathcal G-\mathcal H=E_K-TS+(P_{DIA}-P_{0K})V,
\end{equation}
which consists of three parts, the kinetic energy of thermal atoms, $E_K\sim30$ meV/atom,
the one from entropy, $-TS$,
and the additional work of pressure due to lattice thermal motions, ($P_{DIA}-P_{0K})V$,
where the $P_{0K}$-$V$ EOS is shown in Fig.\ref{fig:4}
and the relative deviations of the EOS, ($|P_{DIA}-P_{0K}|/P_{DIA}$),
are less than about $1\%$ in the pressure range up to $600$ GPa.

The relationship of $\Delta_{thermal}$-$V$ is shown in Fig.\ref{fig:5},
where the total thermal contributions to the FCC phase are overall above $200$ meV/atom
and gets larger at higher pressure zone (HCP and BCC phases).
At the lower pressure zone, $P_{DIA}<200$ GPa,
the pressure difference, $P_{DIA}-P_{0K}$,
corresponds to about $2$ GPa and the atomic volume is about $10$-$14$ {\AA}$^3$/atom,
which leads to $(P_{DIA}-P_{0K})V$ being about $\sim150-190$ meV/atom.
Since the average kinetic energy is on the order of $30$ meV/atom,
the entropy part of the thermal contribution can be estimated to be around the order of $-10$ meV/atom,
which is quite small compared with the other two parts of the thermal contributions.
On the other hand, 
at higher pressure zone,
the contribution of $(P_{DIA}-P_{0K})V$ is estimated to be about $230$ meV/atom, and thus,
the part of entropy contribution becomes larger to be $-30$ meV/atom,
which is on the same order as that of the average kinetic energy.
It is interesting to see that the total thermal contributions increase monotonically up to $\sim350$ meV/atom as the lattice is finally compressed at about $600$ GPa,
which clearly indicates the importance of thermal contributions to the phase stability of crystalline Al under a ultrahigh circumstance at room temperature.
Based on above comparisons, 
the different properties of phase transitions determined by DIA and enthalpy at $0$K cannot be simply attributed to the computational variations but 
a manifested impacts from the thermal contributions.
% with the corresponding phase boundaries locate at $V/V_0=0.55$ for the FCC$\rightarrow$HCP and $V/V_0=0.43$ for the HCP$\rightarrow$BCC
% accompanied with volume changes of $-0.66\%$ and $-0.85\%$ respectively.
\section{Conclusion}
\label{sec:5}
In conclusion,
following the pathway of ensemble theory, 
we apply DIA to solve the PFs of the three phases of crystalline Al,
the derived Gibbs FEs are used as the criterion to investigate the structural phase stability under ultrahigh pressures at room temperature and
the obtained equilibrium structures, transition pressures and EOS are in an excellent agreement with the results from the recent experiment by Dewaele et al\cite{Dewaele2018},
especially validating that the transition pressure of FCC$\rightarrow$HCP is much lower and the one of HCP$\rightarrow$BCC is much higher than those reported in other experiments.
Furthermore,
a comparison with the results obtained by the criterion of enthalpy at $0$K shows that, at room temperature,
the thermal contributions to the Gibbs FE are overall larger than $200$ meV/atom and play a nontrivial role in determining the phase stability, especially at high-pressure extreme conditions.
The high efficiency and precision of DIA to PF make it possible to implement the FE with \emph{ab initio} computational accuracy as the very criterion to study phase behaviors of more condensed-matters materials under extreme conditions in the future.

\section*{Acknowledgement}
The authors are grateful to Y. Akahama for kindly providing the raw experimental data in Ref.\cite{akahama2006}.
Part of the computational tasks was conducted in HPC platform supported by The Major Science and Technology Infrastructure Project of Material Genome Big-science Facilities Platform supported by Municipal Development and Reform Commission of Shenzhen.

\section*{Supplementary Material}
The supplementary material contains the potential energies and potential-energy curves of the FCC, HCP and BCC structures determined by DFT computations,
and the analysis of estimated variations of Gibbs FE due to intrinsic DFT computational errors.

\section*{Data Availability Statement}
The data that supports the findings of this study are available within the article and its supplementary material.

% \section*{Declaration of Competing Interest}
% The authors declare that they have no known competing financial interests or personal relationships that could have appeared to influence the work reported in this paper.
%%       \citet{<label>} ==> Jones et al. [21]
%%       \citep{<label>} ==> [21]
%%

\bibliographystyle{elsarticle-num} 
% \bibliography{ref}
\biboptions{sort&compress}

\end{document}